\documentclass[%
reprint,
groupedaddress,
amsmath,amssymb,
aps,
prl,
superscriptaddress]{revtex4-1}
\usepackage{graphicx}
\usepackage{dcolumn}
\usepackage{bm}
\usepackage{physics}
\usepackage{color}
\RequirePackage[colorlinks=true,linkcolor=blue,urlcolor=blue,hyperfootnotes=false,citecolor=blue]{hyperref}
\usepackage{upgreek}
\begin{document}

\title{Spark-Induced Shockwave Dynamics Revealed via Nonresonant Four-Wave Mixing}

\author{Marios Kounalakis}
\email{marios.kounalakis@gmail.com}

\affiliation{Luxembourg Institute of Science and Technology (LIST), 4362 Esch-sur-Alzette, Luxembourg}
\author{Mikhail N. Shneider}
\affiliation{Department of Mechanical \& Aerospace Engineering, Princeton University, 08544 Princeton, New Jersey, USA}
\affiliation{Luxembourg Institute of Science and Technology (LIST), 4362 Esch-sur-Alzette, Luxembourg}
\author{Alexandros Gerakis}
\email{alexandros.gerakis@list.lu}

\affiliation{Luxembourg Institute of Science and Technology (LIST), 4362 Esch-sur-Alzette, Luxembourg}

\date{\today}

\begin{abstract}
We report on the experimental detection of shockwave dynamics produced in a spark discharge, using a nonresonant four-wave mixing optical technique.
In particular, we observe the spark-induced local density perturbation across a millimeter-range probe volume, centered on the discharge, via single-shot coherent Rayleigh-Brillouin scattering.
We detect the emergence of shock-induced flow velocities, which appear as distinct features in the spectrum, and monitor their dynamic evolution from a few hundred nanoseconds to microseconds after the spark.
Finally, we benchmark our measurements against simulations based on a one-dimensional compressible flow model.
Our results pave the way for quantitative measurements of highly non-uniform transient flows in challenging environments featuring non-equilibrium gas kinetics.
\end{abstract}
\maketitle

\emph{Introduction.}~Non-equilibrium and non-stationary flows describe the dynamic, time-dependent behavior of fluids far from steady-state or equilibrium conditions, encompassing phenomena such as turbulence, shocks, and phase transitions that are only partially understood~\cite{park1989nonequilibrium, Pope_2000}.
Being able to predict and control their behavior is crucial for a wide range of applications in chemical and aerospace engineering, such as thermal protection of supersonic flight and spacecraft re-entry~\cite{laux2007spectroscopic, Gaitonde2015progress, Grisch2002advanced, battat2022nonlinear,zhang2022review, Aiken2025review}, as well as plasma-assisted processes, such as enhanced combustion, pollutant removal, and gasdynamic flow control~\cite{starikovskiy2021gasdynamic, Adamovich2022roadmap, wang2024shock}.
Furthermore, experimentally studying such transient flows, especially in regimes that challenge model predictions, is essential for developing more advanced theories, potentially leading to the discovery of new principles in thermodynamics, fluid dynamics, and plasma physics~\cite{Shneider2006turbulent, Nijdam2020the, Ruggeri1993breakdown,  rezzolla2002new, noronha2022transient, Hafskjold2021theory, Bruggeman2017foundations}.

In this regard, nanosecond repetitively pulsed (NRP) discharges exhibit rich physics, encompassing complex plasma kinetics, non-equilibrium thermodynamics and transient flows~\cite{pai2010transitions, Wang2020nanosecond, starikovskiy2021gasdynamic}.
The \emph{spark} regime, in particular, may exhibit high degrees of ionization and ultrafast gas heating processes that are not fully understood at a fundamental level~\cite{Pai2010nanosecond, Popov2016pulsed, Lo2017streamer, Minesi2020fully, Acciarri2022strong}.
Moreover, as a result of highly localized energy deposition on timescales that are much shorter than the hydrodynamic response of the gas medium, atmospheric spark discharges typically give rise to shockwaves~\cite{Plooster1970shock, Akram1996the, xu2014thermal, dumitrache2019hydrodynamic}.
The spark-induced shock structure is particularly complex and can lead to the formation of boundary layers and recirculation zones, however, a universal description of their dynamics is still lacking in the literature~\cite{Bane2015investigation, Singh2020vortex, adams2023gas, Roger2025origin}.

Experimental efforts to study these phenomena focus primarily on optical imaging, e.g. using shadowgraphy and schlieren~\cite{kogelschatz1972quantitative, xu2014thermal, stepanyan2017large, dumitrache2019hydrodynamic}.
While imaging techniques are useful, they generally do not provide quantitative measurements of the flow field and gas properties, and are limited by line-of-sight averaging and optical interference from the discharge, as well as limited spatiotemporal resolution.
To this end, laser diagnostics are more suitable for flow velocimetry~\cite{Miles2015optical, danehy2015hypersonic, danehy2018non, miles2021localized}.
Focused laser differential interferometry~\cite{Smeets1977flow, Parziale2015observations}, offers high precision, although it is constrained by line-of-sight signal integration, in addition to strong density gradients potentially destroying fringe visibility.
Established techniques such as laser Doppler velocimetry~\cite{Stevenson1982laser} and particle image velocimetry~\cite{adrian2011particle} rely on scattering light from seeded particles in the flow.
In addition, more advanced laser-based diagnostics rely on resonant excitation of particles in the flow and monitoring their fluorescence, such as laser-induced fluorescence and molecular tagging velocimetry~\cite{Hiller1983laser, hanson1988planar, Sanchez2014vibrationally,Leonov:24}.
However, the resonant character of such techniques limits their scope of applications, especially when using seeded gases, which may alter the flow properties and cause instabilities in the extreme conditions surrounding spark discharges.
Rayleigh scattering techniques~\cite{Miles2001laser} offer a nonresonant alternative, being primarily used to probe the translational gas temperature.
The main drawback, as in the other linear techniques, is that the signal is isotropic, therefore, hampered by background noise.
Consequently, direct measurements of the flow field within the spark-induced shock-forming region remain largely inaccessible.

In this Letter, we present a scheme for measuring the shockwave dynamics produced in a spark discharge using nonresonant four-wave mixing (FWM), for the first time to the best of our knowledge.
In particular, we probe the velocity manifold of gas molecules at the center of the discharge and up to $\sim1$~mm in the direction of shock propagation, using single-shot coherent Rayleigh-Brillouin scattering (CRBS)~\cite{gerakis2013single}.
While the technique is species-independent, the choice of $\mathrm{CO}_2$ is motivated by its strong polarizability, high breakdown voltages and reproducible discharges, as well as highly promising environmental and aerospace applications~\cite{pietanza2021advances}.
We observe the effects of the spark on the spectrum, which is partially Doppler-shifted as a result of shock-induced mass displacement.
We monitor the induced flow velocity dynamics in the post-discharge regime, thereby imaging the shockwave evolution in time.
Our findings are supported by simulations of the system dynamics using a one-dimensional compressible flow model with realistic experimental parameters.
Our results enable quantitative studies of flow velocities in highly non-uniform, transient flows, which may lead to a full understanding of the underlying mechanisms governing the gas dynamics in non-equilibrium and non-Maxwellian plasma discharges.

\begin{figure}[t]
  \begin{center}
  \includegraphics[width=1.0\linewidth]{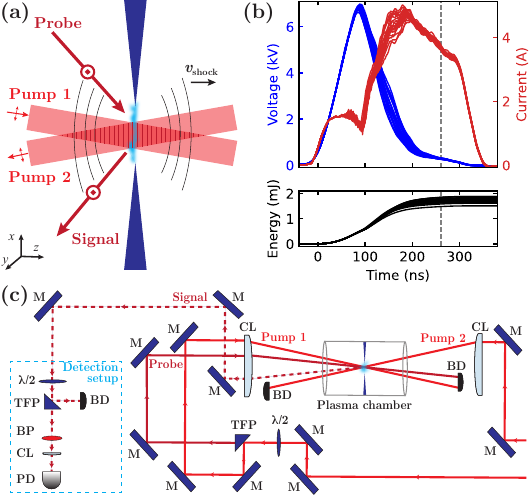}
  \end{center}
  \caption{
  {{Experimental setup.}
  (a)~Schematic illustrating the electrode configuration producing the discharge and the shockwave propagation along the optical lattice.
  The latter is formed by the interference of two equi-polarized pumps in the gas medium.
  The probe (out-of-plane, orthogonally polarized) scatters off the lattice at the Bragg angle and the resulting FWM signal carries information on the velocity of gas particles along $\hat{z}$.
  (b)~Current/voltage profiles during the discharge (top) and deposited energy (bottom), shown for the measurements presented in Fig.~\ref{fig:VDF_evolution_symmetric}.
  The dashed lines indicate when $99~\%$ of the total energy is deposited.
  (c)~Optical setup configuration.
  The signal travels $\sim 3$~m before reaching the detection setup (cyan box).
  M:~plane mirror, $\lambda/2$:~half-wave plate, TFP:~thin-film polarizer, CL:~plano-convex lens, BD:~beam dump, BP:~bandpass filter, PD:~photo-diode.}
  }
  \label{fig:setup}
\end{figure}

\emph{Experimental setup.}~The experimental setup is presented in Fig.~\ref{fig:setup}.
The discharge is generated by applying a high-voltage ($\sim10$~kV) square pulse across two tungsten electrodes with a $100~\upmu\mathrm{m}$ tip radius and $d=1.8$~mm gap, in {$\mathrm{CO}_2$} gas at room temperature and atmospheric pressure.
The voltage and current waveforms, as well as total deposited energy, for the measurements reported here are shown in Fig.~\ref{fig:setup}(b).
In the \emph{spark} regime, the voltage reaches a breakdown around $7$~kV, leading to a $\sim5$~A peak current and total deposited energy of $1.74\pm0.09$~mJ.

We detect the relative density and velocity distribution of particles moving orthogonal to the filament propagation, using single-shot CRBS.
In this FWM scheme, two counter-propagating equi-polarized laser beams ($\lambda\simeq1064$~nm), termed \emph{pump~1} and \emph{pump~2}, are focused and overlap at the center of the discharge, as depicted in Fig.~\ref{fig:setup}(a).
Here, the pumps cross at an angle $\phi/\pi\simeq0.98$, resulting in an interference pattern with a grating wavelength $\lambda_g = \lambda/(2\sin{(\phi/2)})\simeq0.5~\upmu\mathrm{m}$.
Consequently, all neutral gas particles and ions in the probe volume move towards the antinodes of the interference pattern (where the electric field is maximum), forming an optical lattice.
This is a result of the optical dipole force, $F_D=-\alpha_\mathrm{eff}q|E_1E_2|\sin{\left(qz-2\pi\Delta f~t\right)}$, acting on particles with nonresonant effective polarizability $\alpha_\mathrm{eff}$, where $E_i$ is the complex electric field of each pump, $q=|\vec{k}_1-\vec{k}_2|=2\pi/\lambda_g$ is the grating wavevector, and $\Delta f$ is the frequency difference of the pumps~\cite{boyd2020nonlinear}.
A third beam, termed \emph{probe}, is scattered off the optical lattice at the Bragg angle, $\theta=\arcsin{(\lambda_\mathrm{probe}/2\lambda_g)}$, resulting in a coherent \emph{signal} beam in the direction determined by the phase matching condition, $\vec{k}_{\mathrm{signal}}=\vec{k}_\mathrm{probe}+\vec{k}_1-\vec{k}_2$, with intensity $I_{0}\propto \Delta\rho^2I_1 I_2 I_\mathrm{probe}$, where $\Delta\rho$ is the gas density perturbation along the probe volume length,  $L$~\cite{pan2002coherent}.

When $\Delta f=0$, the optical lattice is stationary in time and the signal is the result of coherent Rayleigh scattering from particles with velocity $v\simeq0$~\cite{grinstead2000coherent}.
When a relative frequency difference is introduced between the pumps, the grating moves with phase velocity $v_\mathrm{ph} = \frac{\Delta f~\lambda_g}{2\sin{(\phi/2)}}$.
In this case, $F_D$ acts on particles moving with $v\simeq v_\mathrm{ph}$ and the probe is coherently scattered off this lattice.
By measuring $I_{0}$ as a function of $\Delta f$ one can retrieve the CRBS spectrum, which carries information on the velocity distribution of particles along $\hat{z}$~\cite{pan2002coherent, gerakis2025seedless}.
We implement this in a single-shot measurement by chirping one of the pumps, as detailed in Ref.~\cite{gerakis2013single}.

The optical setup is schematically depicted in Fig.~\ref{fig:setup}(c).
Both pumps and the probe are generated by a custom-built laser system that is capable of producing flat-top pulses of $\sim100$~ns with energy $\sim2$~J per pulse, at $5$~Hz repetition rate~\cite{Bak:22, karatodorov2025high}.
Pump~2 is chirped at a rate $\sim14~\mathrm{MHz/ns}$.
The signal is routed $\sim3$~m away from the discharge to reduce background optical noise.
The detection setup consists of a $\lambda/2$-TFP combination, to suppress stray light from the pumps, followed by a $1064$~nm bandpass filter and a plano-convex lens focusing the signal to an InGaAs photodiode detector.

\begin{figure}[t]
  \includegraphics[width=1\linewidth]{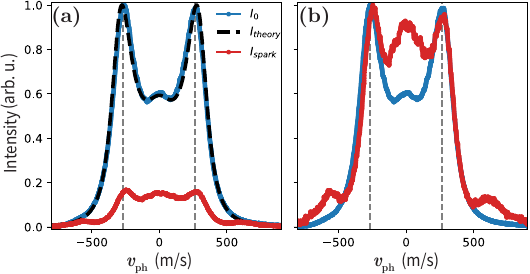}
  \caption{
  {Detection of the spark.}
  (a)~CRBS spectra measured in quiescent (no discharge) conditions ($I_0$, blue), and $\sim0.6~\upmu\mathrm{s}$ after the spark ($I_\mathrm{spark}$, red) in room-temperature, atmospheric {$\mathrm{CO}_2$} gas.
  Both plots are normalized by $I_0^\mathrm{MAX}$.
  The dashed black curve ($I_\mathrm{theory}$) shows the theoretically predicted spectrum for the quiescent case.
  The vertical dashed lines indicate the {$\mathrm{CO}_2$} speed of sound.
  (b)~Same spectra, each normalized by their corresponding maximum signal intensities, showcasing drastically different gas kinetics in the spark regime, highlighted by supersonic flow velocities.
  }
  \label{fig:CRBS_shockwave}
\end{figure}

\emph{Detection scheme.}~A typical CRBS spectrum reflecting the velocity distribution of {$\mathrm{CO}_2$} gas particles in quiescent conditions, i.e. when there is no discharge, is plotted by the blue lineshape in Fig.~\ref{fig:CRBS_shockwave}(a).
The black dashed curve plots the theoretical prediction for the same thermodynamic conditions ($T_0=297$~K, $p_0=0.95$~bar) using the analytical calculations from Pan's S7 kinetic model~\cite{pan2004coherent}.
In addition to the Rayleigh peak, two Brillouin peaks are obtained, as a result of light scattering off particles moving at $\pm v_\mathrm{sound}$ (vertical dashed lines) along the lattice.

The red solid curve in Fig.~\ref{fig:CRBS_shockwave}(a) plots the CRBS spectrum measured at the center of the inter-electrode gap $\sim0.6~\upmu\mathrm{s}$ after the spark.
The plotted spectra are averaged over $5$ single-shot measurements to reduce background electrical noise from the discharge.
Both spectra are normalized by $I_0^\mathrm{MAX}$, illustrating the difference between the observed signal intensities.
The relative density ratio in the probe volume is found to be $N_\mathrm{spark}/N_0=\sqrt{\left(\int{dv I_\mathrm{spark}}\right)/\left(\int{dv I_0}\right)}\sim 45~\%$.
Note that this does not reflect the total degree of ionization in the discharge, since the probe volume length, $L\simeq2~$mm, is much larger than the channel radius, $r_\mathrm{ch}\simeq100~\upmu$m.

\begin{figure*}[t]
  \includegraphics[width=0.8\linewidth]{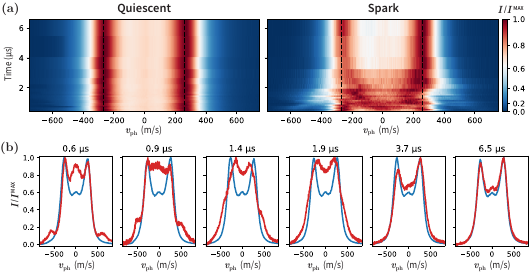}
  \caption{
  {Shockwave dynamics.}
  (a)~Right: Heatmap plotting the evolution of the CRBS spectrum in {$\mathrm{CO}_2$} after the spark.
  The dashed lines indicate $|v_\mathrm{sound}|$ in quiescent conditions.
  The probe volume is centered at the discharge channel (see Fig.~\ref{fig:setup}(a)).
  Left: Same plots measured in quiescent (no spark) conditions.
  (b)~Selected normalized spectra from (a) highlighting the main differences between quiescent (blue) and spark (red) conditions over time.
  }
  \label{fig:VDF_evolution_symmetric}
\end{figure*}

\begin{figure}[t]
  \includegraphics[width=1\linewidth]{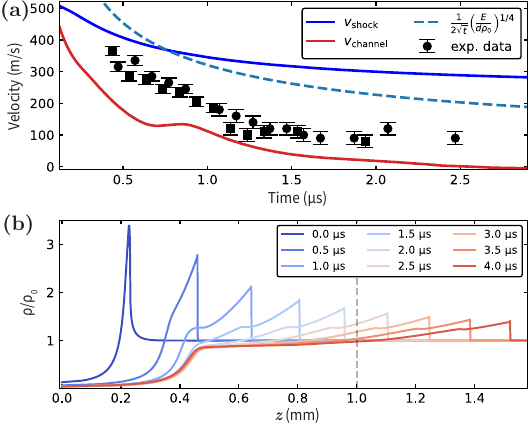}
  \caption{Evolution of shock-induced flow velocity.
  (a)~Experimentally obtained flow velocities as a function of time in the spark regime, extracted from the Doppler-shifted spectra with the probe volume centered (circles) and offset (squares) with respect to the discharge.
  The blue and red curves plot the shock front and heated channel velocities, respectively, estimated from the gas dynamics model; the dashed curve plots the estimated shock speed from similarity analysis (see text for details).
  (b)~Evolution of the relative density in the shock propagation direction. The dashed line indicates the approximate probe volume length.
  }
  \label{fig:SimulationsData}
\end{figure}

The influence of the spark discharge on the gas velocity distribution is showcased in Fig.~\ref{fig:CRBS_shockwave}(b), which plots the normalized spectra $I_0/I_0^\mathrm{MAX}$ (blue) and $I_\mathrm{spark}/I_\mathrm{spark}^\mathrm{MAX}$ (red).
Interestingly, after the spark, Rayleigh scattering appears to be more pronounced relative to the Brillouin peak intensities.
Furthermore, two additional peaks appear at $v\simeq\pm600$~m/s, indicating a spark-induced flow at supersonic speeds.
Note that since $L\gg r_\mathrm{ch}$, the observed signal is dominated by scattering off gas particles outside the heated channel.
We expect an outward propagating shockwave perpendicular to the discharge channel, as reported with schlieren imaging in similar setups~\cite{stepanyan2017large, dumitrache2019hydrodynamic}.
Since the width of the perturbed area is much smaller than $L$ (see also Fig.~\ref{fig:SimulationsData}), we observe a partially \emph{Doppler-shifted} spectrum~\cite{gerakis2025seedless, kumar2025non}.
More specifically, the ``outward-shifted'' peaks result from light scattering off particles moving at $ \pm(v^\prime+v_\mathrm{sound})$, where $ v^\prime$ is the shock-induced mass displacement speed.

\emph{Imaging shockwave dynamics.}~We image the evolution of the propagating shockwave via the CRBS spectrum, by introducing a relative time delay between the laser pulses and the high-voltage pulse.
The measured spectra as a function of time, are plotted in Fig.~\ref{fig:VDF_evolution_symmetric}.
Here, $t=0$ is defined as the time when $99~\%$ of the total energy, from the high-voltage pulse, is deposited into the plasma (dashed line in Fig.~\ref{fig:setup}(b)).
The spark and quiescent spectra are acquired sequentially, by turning on/off the discharge, at each time step.
For each spectrum, we have also recorded and subtracted the remaining optical background ($\lesssim5\%$) when the probe is blocked.

The dynamic evolution of the CRBS spectra, presented in Fig.~\ref{fig:VDF_evolution_symmetric}, can be categorized into three regimes of post-spark flow dynamics.
First, up to $\sim1~\upmu\mathrm{s}$, the obtained spectra feature more pronounced Rayleigh peaks (relative to the Brillouin peaks) and two distinct peaks at $|v|> v_\mathrm{sound}$.
As discussed, these peaks are essentially the ``outward-shifted" Brillouin peaks at $\pm(v_\mathrm{sound}+v^\prime(t))$.
An additional pair of ``inward-shifted" peaks is expected around $\pm(v_\mathrm{sound}-v^\prime(t))$, and since $v_\mathrm{sound}\simeq v^\prime(t)$ at this timescale, they contribute to the more pronounced Rayleigh peak that is observed.

The second regime, $1~\upmu\mathrm{s}\lesssim t\lesssim 3~\upmu\mathrm{s}$, is characterized by drastically different spectral features, exhibiting diminished Rayleigh and distorted Brillouin peaks, as a result of the induced Doppler shifts.
Moreover, the ``outward-shifted'' peaks begin to move towards $v_\mathrm{sound}$, showcasing a decelerating shockwave.
In addition, as explained in the next section, the moving density perturbation begins to spread and covers a larger part of the probe volume (Fig.~\ref{fig:SimulationsData}(b)), leading to more prevalent ``inward-shifted'' peaks appearing at $\pm(v_\mathrm{sound}-v^\prime(t))$.

Finally, for $3~\upmu\mathrm{s}\lesssim t\lesssim 6~\upmu\mathrm{s}$, a more equilibrium-like CRBS spectrum is obtained with a net flow at $\sim v_\mathrm{sound}$ remaining in the probe volume, as indicated by the asymmetry in the two Brillouin peaks.
This asymmetry is a consequence of the probe volume not being perfectly centered in the discharge channel.
Finally, for $t> 6~\upmu\mathrm{s}$, the velocity distribution in the spark regime returns to normal, as the shock-induced density perturbation has left the probe volume (see also Fig.~\ref{fig:SimulationsData}(b)).

\begin{figure}[t]
  \includegraphics[width=1\linewidth]{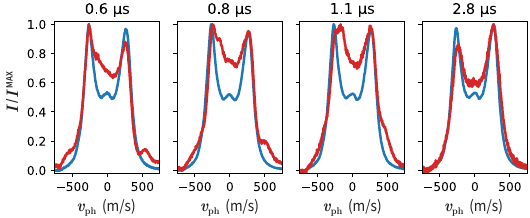}
  \caption{
  Shockwave evolution revealed in the CRBS spectrum with the electrodes offset such that the probe volume lies mostly on one side with respect to the discharge center.
  Similar dynamics as in Fig.~\ref{fig:VDF_evolution_symmetric} are observed, however, primarily positive shock-induced velocities are detected.
  }
  \label{fig:VDF_evolution_asymmetric}
\end{figure}

The spectra shown in Fig.~\ref{fig:VDF_evolution_symmetric} are measured with the probe volume centered around the core of the discharge channel.
We additionally perform the same measurements with the electrode locations shifted, such that the probe volume is offset towards the right side of the optical lattice, thereby capturing the shockwave propagation primarily at positive velocities~(see Fig.~\ref{fig:setup}(a)).
Selected spectra are presented in Fig.~\ref{fig:VDF_evolution_asymmetric}, further supporting the dynamics presented above.
As expected, in this configuration we only observe the ``outward-shifted'' right Brillouin peak appearing at $ v_\mathrm{sound}+v^\prime(t)$ and the ``inward-shifted'' left peak at $-v_\mathrm{sound}+v^\prime(t)$.

\emph{Gas dynamics benchmarking.}~We benchmark the experimental findings using one-dimensional gas simulations in cylindrical coordinates.
The model simulates a radially symmetric, compressible flow in $\mathrm{CO}_2$ gas, using the standard formulation in Lagrange's mass coordinates~\cite{brode1955numerical, wilkins1980use, xu2014thermal}.
Our model additionally includes a temperature–dependent specific heat ratio and thermal conductivity for $\mathrm{CO}_2$~\cite{huber2016reference}.
The discharge channel is assumed to be uniform between the electrodes, while the current density across the discharge is modeled using a Gaussian profile with a channel radius, $r_\mathrm{ch}=100~\upmu\mathrm{m}$.
The experimentally obtained current-voltage profiles presented in Fig.~\ref{fig:setup}(b) are used as input for the deposited energy, scaled by an effective heat factor ($\sim40~\%$) that is determined by the size of $r_\mathrm{ch}$~\cite{lin1954cylindrical}.

The simulation results along with the experimental data are plotted in Fig.~\ref{fig:SimulationsData}(a).
Each data point represents $v^\prime(t)$, which is extracted by monitoring the Doppler-shifted peaks of each spectrum for both centered (circles) and offset (squares) cases.
The blue curve plots the velocity of the shockwave front, $v_\mathrm{shock}$, calculated by monitoring the evolution of the sharp density gradient profile, shown in Fig.~\ref{fig:SimulationsData}(b).
The red curve plots the velocity of the heated channel, $v_\mathrm{channel}$, defined by the condition, $\rho/\rho_0=0.5$, for the density ratio.
The dashed curve plots the analytical expression for strong cylindrical shocks from similarity analysis, $(2\sqrt{t})^{-1}\left({E}/{d\rho_0}\right)^{1/4}$, assuming an instantaneous energy release $E$~\cite{lin1954cylindrical,sedov2018similarity}.

In the initial phase, we observe that $v^\prime(t)$ is closer to $v_\mathrm{shock}$.
During this timescale, the propagating density gradient is mostly confined in space, as shown in Fig.~\ref{fig:SimulationsData}(b).
However, at later times the density profile is more stretched out in space, with each component moving at a different velocity.
The measured $v^\prime(t)$ at this timescale is therefore the average of a range of velocities contributing to the Doppler-shifted peaks.
The dashed line in Fig.~\ref{fig:SimulationsData}(b) indicates the probe volume length beyond which the shock-induced flow is no longer detected.

\emph{Summary-Outlook.}~In summary, we have experimentally measured the shockwave dynamics produced in a spark discharge, via a seedless and non-intrusive FWM optical technique.
To the best of our knowledge, this is the first demonstration of direct velocity measurements in a highly non-uniform transient flow using nonresonant, nonlinear optics.
We have experimentally characterized the single-shot CRBS spectrum in the post-discharge regime, from $~\sim400$~ns up to a few microseconds, probing at two different positions: one centered and one offset with respect to the discharge channel.
By monitoring the Doppler-shifted spectrum as a function of time, we demonstrate that the average flow velocity of the shock-induced density perturbation can be extracted.
We benchmark our measurements against simulations using a one-dimensional model of compressible flow and analytical predictions for cylindrical shockwaves.

Our results open up new opportunities for fundamental research, enabling remote and non-intrusive experimental studies in plasma dynamics, fluid mechanics and thermodynamics.
In terms of applications, they enable more quantitative measurements of flow dynamics in $\mathrm{CO}_2$ spark discharges, which are highly relevant for environmental remediation, e.g. plasma-assisted decarbonization, as well as for aerospace and planetary plasma studies~\cite{Guerra2022plasmas}.
Future work, could lead to a full reconstruction of the flow velocity field in the discharge as well as local density and temperature measurements, by combining different CRBS beam geometries, e.g. co-propagating~\cite{Gerakis2011coherent}, and multi-point measurement schemes~\cite{kumar2025multi, kumar2025simultaneous}.
Such experiments could shed light on fundamental questions, e.g. concerning the physical mechanism responsible for ultrafast heating in thermal and non-thermal sparks, without relying on Maxwellian distributions or local thermal equilibrium~\cite{Minesi2020fully, Acciarri2022strong}.

\emph{Acknowledgments.}
We acknowledge technical support by O.~Bouton, M.~Gerard, M.~Kummel and N.~Cremonesi.
We thank M.~Simeni Simeni, G.~Trayner, A.U.~Kumar for discussions and A.~Hosseinnia for reading and commenting on the manuscript.
We acknowledge financial support by the Luxembourg National Research Fund $17838565$ (ULTRAION), $19525000$ (ETALON) and $15480342$ (FRAGOLA).

\bibliography{BibMarios}

\end{document}